\RequirePackage[OT1]{fontenc}
\documentclass[journal]{IEEEtran}

\usepackage{amsmath,amsfonts}
\usepackage{graphicx}
\usepackage[noadjust]{cite}
\usepackage{makecell}
\usepackage{multirow}
\usepackage[acronym]{glossaries}
\usepackage{svg}
\usepackage{url}
\usepackage{algorithm}
\usepackage{algpseudocode}
\usepackage[column=O]{cellspace}
\usepackage{siunitx}
\usepackage{silence}
\usepackage{hyperref}
\usepackage{microtype}
\usepackage{booktabs}
\usepackage{orcidlink}

\ifCLASSOPTIONcompsoc
  \usepackage[caption=false,font=normalsize,labelfont=sf,textfont=sf]{subfig}
\else
  \usepackage[caption=false,font=footnotesize]{subfig}
\fi

\makeatletter
\newenvironment{breakablealgorithm}
  {%
   \begin{center}
     \refstepcounter{algorithm}%
     \hrule height.8pt depth0pt \kern2pt%
     \renewcommand{\caption}[2][\relax]{%
       {\raggedright\textbf{\ALG@name~\thealgorithm} ##2\par}%
       \ifx\relax##1\relax %
         \addcontentsline{loa}{algorithm}{\protect\numberline{\thealgorithm}##2}%
       \else %
         \addcontentsline{loa}{algorithm}{\protect\numberline{\thealgorithm}##1}%
       \fi
       \kern2pt\hrule\kern2pt
     }
  }{%
     \kern2pt\hrule\relax%
   \end{center}
  }
\makeatother

\makeatletter
\def\bstctlcite{\@ifnextchar[{\@bstctlcite}{\@bstctlcite[@auxout]}}
\def\@bstctlcite[#1]#2{\@bsphack
  \@for\@citeb:=#2\do{%
    \edef\@citeb{\expandafter\@firstofone\@citeb}%
    \if@filesw\immediate\write\csname #1\endcsname{\string\citation{\@citeb}}\fi}%
  \@esphack}
\makeatother

\newacronym{2d}{2D}{two-dimensional}

\newacronym{brir}{BRIR}{binaural room impulse response}

\newacronym{dnn}{DNN}{deep neural network}
\newacronym{ffnn}{FFNN}{feedforward neural network}

\newacronym{ic}{IC}{interaural coherence}
\newacronym{ild}{ILD}{interaural level difference}
\newacronym{irm}{IRM}{ideal ratio mask}
\newacronym{itd}{ITD}{interaural time difference}

\newacronym{logfbe}{log-FBE}{log-filterbank energy}

\newacronym{mse}{MSE}{mean squared error}

\newacronym{pesq}{PESQ}{perceptual evaluation of speech quality}

\newacronym{relu}{ReLU}{rectified linear unit}
\newacronym{rms}{RMS}{root mean square}

\newacronym{snr}{SNR}{signal-to-noise ratio}
\newacronym{sisnr}{SI-SNR}{scale-invariant signal-to-noise ratio}
\newacronym{stoi}{STOI}{short-term objective intelligibility}
\newacronym{estoi}{ESTOI}{extended short-term objective intelligibility}

\newacronym{tf}{TF}{time-frequency}
\newacronym{stft}{STFT}{short-time Fourier transform}

\newacronym{svm}{SVM}{support vector machine}

\newacronym{asr}{ASR}{automatic speech recognition}
\newacronym{sid}{SID}{speaker identification}

\newacronym{drr}{DRR}{direct-to-reverberant energy ratio}

\newacronym{lstm}{LSTM}{long short-term memory}

\newacronym{iid}{IID}{independent and identically distributed}
\newacronym{ood}{OOD}{out-of-distribution}
\newacronym{dg}{DG}{domain generalization}

\newcommand\narrowstyle{\SetTracking{encoding=*}{-50}\lsstyle}

\def\dataset{\mathcal{D}}
\def\dtrain{\dataset^{\text{train}}}
\def\dtrainref{\dataset^{\text{train}}_{\text{ref}}}
\def\dtest{\dataset^{\text{test}}}
\def\dtraini{\dataset^{\text{train}}_i}
\def\dtrainrefi{\dataset^{\text{train}}_{\text{ref}, i}}
\def\dtesti{\dataset^{\text{test}}_i}

\def\cond{C}
\def\condtrain{\cond^{\text{train}}}
\def\condtest{\cond^{\text{test}}}

\def\corpus{\mathcal{C}}
\newcommand{\corp}[2]{\corpus_{\text{#1}, #2}}
\newcommand{\corpi}[1]{\corpus_{#1}}

\newcommand{\corptraini}[1]{\corpus^\text{train}_{#1}}
\newcommand{\corptesti}[1]{\corpus^\text{test}_{#1}}
\def\nspeech{N_\text{speech}}
\def\nnoise{N_\text{noise}}
\def\nroom{N_\text{room}}

\def\set{\mathcal{S}}
\def\sspeech{\set_\text{speech}}
\def\snoise{\set_\text{noise}}
\def\sroom{\set_\text{room}}
\def\strainspeech{\set^\text{train}_\text{speech}}
\def\strainnoise{\set^\text{train}_\text{noise}}
\def\strainroom{\set^\text{train}_\text{room}}
\def\stestspeech{\set^\text{test}_\text{speech}}
\def\stestnoise{\set^\text{test}_\text{noise}}
\def\stestroom{\set^\text{test}_\text{room}}
\def\strainspeechi{\set^\text{train}_{\text{speech}, i}}
\def\strainnoisei{\set^\text{train}_{\text{noise}, i}}
\def\strainroomi{\set^\text{train}_{\text{room}, i}}
\def\stestspeechi{\set^\text{test}_{\text{speech}, i}}
\def\stestnoisei{\set^\text{test}_{\text{noise}, i}}
\def\stestroomi{\set^\text{test}_{\text{room}, i}}
\def\straindi{\set^\text{train}_{d, i}}
\def\stestdi{\set^\text{test}_{d, i}}

\def\model{\mathcal{M}}
\def\modelref{\model_{\text{ref}}}
\def\modeli{\model_{i}}
\def\modelrefi{\model_{\text{ref}, i}}

\def\score{E}
\def\scorei{\score_{i}}
\def\scorerefi{\score_{\text{ref}, i}}

\def\dpesq{\Delta\text{\gls{pesq}}}

\def\destoi{\Delta\text{\gls{estoi}}}
\def\dsnr{\Delta\text{\gls{snr}}}

\def\gengap{G_\score}
\def\gdpesq{G_{\dpesq}}

\def\gdestoi{G_{\destoi}}
\def\gdsnr{G_{\dsnr}}

\title{Assessing the Generalization Gap of Learning-Based Speech Enhancement Systems in Noisy and Reverberant Environments}

\author{Philippe~Gonzalez\,\orcidlink{0009-0006-4965-3514},~\IEEEmembership{Student~Member,~IEEE}, Tommy~Sonne~Alstr{\o}m\,\orcidlink{0000-0003-0941-3146},~\IEEEmembership{Member,~IEEE}, Tobias~May\,\orcidlink{0000-0002-5463-5509}}

\begin{document}

\bstctlcite{IEEEexample:BSTcontrol}

\maketitle

\begin{tikzpicture}[remember picture,overlay]
\node[anchor=south,yshift=3pt] at (current page.south) {
  \fbox{\parbox{\dimexpr\textwidth - 2\fboxsep}{
    \footnotesize \copyright 2023 IEEE. Personal use of this material is permitted. Permission from IEEE must be obtained for all other uses, in any current or future media, including reprinting/republishing this material for advertising or promotional purposes, creating new collective works, for resale or redistribution to servers or lists, or reuse of any copyrighted component of this work in other works.
  }}
};
\end{tikzpicture}

\begin{abstract}
The acoustic variability of noisy and reverberant speech mixtures is influenced by multiple factors, such as the spectro-temporal characteristics of the target speaker and the interfering noise, the \gls{snr} and the room characteristics.
This large variability poses a major challenge for learning-based speech enhancement systems, since a mismatch between the training and testing conditions can substantially reduce the performance of the system.
Generalization to unseen conditions is typically assessed by testing the system with a new speech, noise or \gls{brir} database different from the one used during training.
However, the difficulty of the speech enhancement task can change across databases, which can substantially influence the results.
The present study introduces a generalization assessment framework that uses a reference model trained on the test condition, such that it can be used as a proxy for the difficulty of the test condition.
This allows to disentangle the effect of the change in task difficulty from the effect of dealing with new data, and thus to define a new measure of generalization performance termed the generalization gap.
The procedure is repeated in a cross-validation fashion by cycling through multiple speech, noise, and \gls{brir} databases to accurately estimate the generalization gap.
The proposed framework is applied to evaluate the generalization potential of a \gls{ffnn}, Conv-TasNet, DCCRN and MANNER.
We find that for all models, the performance degrades the most in speech mismatches, while good noise and room generalization can be achieved by training on multiple databases.
Moreover, while recent models show higher performance in matched conditions, their performance substantially decreases in mismatched conditions and can become inferior to that of the \gls{ffnn}-based system.
\end{abstract}


\section{Introduction}
\label{sec:intro}

\IEEEPARstart{A}{major} challenge of hearing-impaired listeners is to focus on a specific talker in noisy and reverberant environments.
The presence of noise sources in the acoustic scene can degrade speech intelligibility through spectral masking~\cite{french1947factors,gelfand2017hearing}, while late reflections from reverberation can also deteriorate speech intelligibility of both normal-hearing and hearing-impaired listeners due to temporal smearing~\cite{moncur1967binaural,nabelek1982monaural,kokkinakis2011channel,hazrati2013blind}.
In addition, it was shown that the combined effect of noise and reverberation reduces speech intelligibility to a greater extent than individually~\cite{nabelek1981effect,hazrati2012combined}.
The presence of reverberation also deteriorates the performance of technical applications such as automatic speech recognition~\cite{palomaki2004techniques} and speaker identification~\cite{may2012binaural} systems.
Thus, the development of technical solutions that can enhance speech in noisy and reverberant conditions is of critical importance for a wide range of applications, such as communication devices and hearing aids.

In recent years, deep learning techniques have seen an increasing interest in the speech processing community due to their superior performance over traditional speech enhancement approaches~\cite{wang2012cocktail,xu2015regression,wang2018supervised,bentsen2018benefit}.
However, due to the large number of trainable parameters, the performance of neural network-based systems typically decreases in acoustic conditions that were not included in the training stage~\cite{may2014requirements,chen2017long,pandey2020cross}.
While this is a well-known problem in deep learning, it is particularly prominent in speech processing due to the plethora of parameters affecting the mixture at the input of the system.
These parameters include the spectro-temporal characteristics and the positions of the target speaker and the noise sources in the acoustic scene, the room characteristics such as the reverberation time and the direct-to-reverberant energy ratio, the \gls{snr}, the mixture level and the language.
As a consequence, the variability of noisy and reverberant mixtures is extremely high, which makes it challenging to develop robust learning-based systems that perform well in a wide range of conditions.

Several studies have addressed the generalization problem in speech enhancement~\cite{han2012towards,wang2013towards,wang2015deep,kolbaek2016speech,chen2016large,chen2017long,pandey2020cross,pandey2020learning,healy2021effectively}.
The typical approach to measure the generalization performance consists in testing the system with mixtures created from a speech corpus or a noise database that differs from the one used during training.
For example in~\cite{wang2013towards}, the speech and noise generalization of a \gls{ffnn} combined with a support vector machine was evaluated.
The system was trained with TIMIT~\cite{garofolo1993timit} sentences and 100 environmental noises~\cite{hu2010nonspeech}, and tested with unseen TIMIT and IEEE sentences~\cite{ieee1969ieee}, as well as noises compiled from CHiME~\cite{christensen2010chime}, NOISEX~\cite{varga1993noisex} and the tandem algorithm~\cite{hu2010tandem}.
Similarly in~\cite{kolbaek2016speech}, the generalization of a \gls{ffnn}-based system to unseen speech, noises and \glspl{snr} was evaluated using different combinations of noise types and speaker groups for training and testing.
The noise types consisted of speech babble, speech-shaped noise, recordings from CHiME3~\cite{barker2015third} and sound effects from soundbible.com, while speech utterances were sourced from the phonetically balanced Danish speech corpus ``\textit{Akustiske Databaser for Dansk}''~\cite{nordisk2010akustiske}.
Both studies showed that training on more diverse acoustic conditions improved generalization performance, especially to unseen \glspl{snr} and speech.
Finally in~\cite{pandey2020cross}, cross-corpus generalization to unseen speech was addressed by using utterances from WSJ~\cite{paul1992design}, VoxCeleb2~\cite{chung2018voxceleb2} and LibriSpeech~\cite{panayotov2015librispeech} for training, and WSJ, TIMIT and IEEE for testing.
The study showed that using a diverse and crowd-sourced corpus for training, such as LibriSpeech, significantly improved the generalization performance of a \gls{lstm} network compared to WSJ, which was recorded in controlled environments.

While testing the system on an unseen database can provide some insight into the generalization abilities, the result depends on the selected databases and would likely be different if another unseen database was used.
Indeed, when using a new database for testing, the model not only has to deal with unseen data, but the difficulty of the speech enhancement task can also change.
For example, noise segments from a database consisting of stationary environmental recordings might be easier to suppress compared to noises with spectro-temporal fluctuations similar to speech.
Similarly, speech corpora recorded in controlled environments can be easier to enhance compared to corpora of crowd-sourced recordings, where the noise level and the recording equipment can change across recordings.
Thus, even if the system is generalizing well, a drop in performance between the training and testing conditions can be measured.
Previous studies have not controlled for the potential change in the difficulty of the task when using different databases, and have not disentangled it from the effect of the model dealing with unseen data.
Additionally, to reduce the sensitivity of the estimated generalization performance to the selected databases, multiple databases should be considered for training and testing in a cross-validation manner.
Finally, the aforementioned studies only considered single-channel mixtures without reverberation, and did not study the generalization to unseen rooms.
In acoustic scenes with a binaural receiver, the acoustic properties of the room and the spatial location of the different sources in the scene can also greatly affect the acoustic mixture at the input of the system.
A mismatch in terms of room and source position between training and testing can thus also lead to a substantial drop in performance.

The present work introduces a systematic framework to accurately assess the generalization performance of learning-based speech enhancement systems.
In order to disentangle the effect of dealing with unseen data from the potential change in task difficulty when testing with new databases, a reference model is trained on the testing condition to provide an upper performance limit.
We define a new measure of generalization performance, termed the generalization gap, as the relative difference between the evaluated model and the reference model in terms of frequently-used objective metrics.
Moreover, to further reduce the influence of individual speech, noise and \gls{brir} databases, the evaluation is repeated in a cross-validation fashion by cycling through multiple combinations of databases for training and testing.
The generalization gap is then averaged across folds for a more accurate estimation.
We use this framework to evaluate the influence of the speech, noise and room dimensions on the generalization performance of four speech enhancement systems: a \gls{ffnn}-based system, Conv-TasNet~\cite{luo2019conv}, DCCRN~\cite{hu2020dccrn} and MANNER~\cite{park2022manner}.
Combined mismatches along multiple dimensions are also investigated.

\section{Generalization performance assessment}
\label{sec:method}

\subsection{Matched and mismatched conditions}

The following section defines an acoustic condition and a dataset in the context of generalization assessment.
In the present framework, a condition $C$ is defined by the set of speech corpora $\sspeech$, the set of noise databases $\snoise$ and the set of \gls{brir} databases $\sroom$ used to generate datasets of noisy and reverberant mixtures.
A condition can be seen as a parametrization of the random process used to generate noisy and reverberant mixtures $X(t)$,
\begin{equation}
    X(t) \hookrightarrow C(\sspeech, \snoise, \sroom),
\end{equation}
with
\begin{align}
    \sspeech &= \{\corp{speech}{1}, \ldots, \corp{speech}{\nspeech}\},\\
    \snoise &= \{\corp{noise}{1}, \ldots, \corp{noise}{\nnoise}\},\\
    \sroom &= \{\corp{room}{1}, \ldots, \corp{room}{\nroom}\},
\end{align}
where $\corp{speech}{1}, \ldots, \corp{speech}{\nspeech}$ are $\nspeech$ different speech corpora, $\corp{noise}{1}, \ldots, \corp{noise}{\nnoise}$ are $\nnoise$ different noise databases, and $\corp{room}{1}, \ldots, \corp{room}{\nroom}$ are $\nroom$ different \gls{brir} databases.
While a condition can be parametrized using other acoustic parameters such as the range of \glspl{snr}, the position of the sources or the mixture level, in this study we only investigate the speech, noise and room effects, and other acoustic parameters vary within the same range across all conditions as described in Sect.~\ref{sec:mixgen}.
With the proposed definition for a condition in mind, a dataset $\dataset$ thus consists of a set of realizations of noisy and reverberant mixtures $X(t)$,
\begin{equation}
    \dataset = \{x_1(t), \ldots, x_{N_x}(t)\},
\end{equation}
where $N_x$ is the number of mixtures in the dataset.

Consider now two datasets $\dtrain$ and $\dtest$ of realizations from the conditions $C^\text{train}$ and $C^\text{test}$ respectively.
We refer to $C^\text{train}$ and $C^\text{test}$ as matched if and only if $\strainspeech, \strainnoise, \strainroom=\stestspeech, \stestnoise, \stestroom$.
If any of the sets of databases used to generate the datasets $\dtrain$ and $\dtest$ are different, then the conditions are denoted as mismatched.
Note that even in the case of matched conditions, we ensure that no speech utterance, noise segment nor \gls{brir} appears in both $\dtrain$ and $\dtest$ (that is, even if $C^\text{train}$ and $C^\text{test}$ are the same, we ensure during the random mixture generation process that this constraint is fulfilled).

A mismatch can occur along one or multiple acoustic dimensions.
For example, if $\strainspeech \neq \stestspeech$ but $\strainnoise = \stestnoise$ and $\strainroom = \stestroom$, then we have a single mismatch.
Similarly if $\strainspeech \neq \stestspeech$, $\strainnoise \neq \stestnoise$ and $\strainroom = \stestroom$, then we have a double mismatch.
Finally if $\strainspeech \neq \stestspeech$, $\strainnoise \neq \stestnoise$ and $\strainroom \neq \stestroom$, then we have a triple mismatch.
In the following, mismatches along a single, two or all three dimensions are investigated.

\subsection{Reference model}\label{sec:ref_model}

\begin{figure}
    \scriptsize
    \centering
    \includeinkscape[width=.95\linewidth]{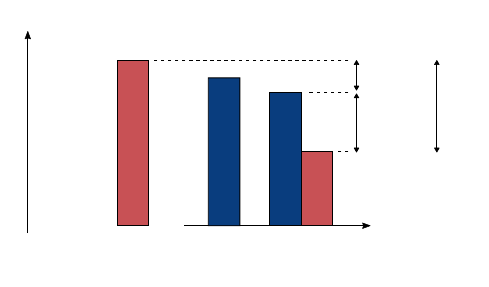}
    \caption{Illustration of how the reference model allows to disentangle the effect of the model observing new data from the change in difficulty of the speech enhancement task.
    The evaluated model $\model$ is trained using a dataset $\dtrain$ from a condition $\condtrain$.
    $\model$ can be evaluated on a dataset $\dtest_\text{match}$ from the same condition, however this would not inform on generalization.
    Instead, $\model$ is evaluated on a dataset $\dtest$ from a new condition $\condtest$.
    Since $\condtrain$ and $\condtest$ are different, the drop in performance is a biased measure of generalization, as this can be influenced by a change in difficulty.
    To overcome this, a reference model $\modelref$ is trained on a dataset $\dtrain_\text{ref}$ from $\condtest$ and evaluated on $\dtest$.~The
    difficulty~of~$\condtest$~should~be~reflected~by~the~performance~of~$\modelref$~on~$\dtest$.}
    \label{fig:gengap}
\end{figure}

When evaluating a learning-based model in a new condition, the performance of the model can change not only due to the fact that the model sees data from a distribution that is different from the one seen during training, but also because the speech enhancement task in the new condition might be inherently more or less difficult.
As a consequence, directly using the performance drop between training and testing as a measure of generalization performance can be misleading.
For example, a very robust model can show poor performance when evaluated with noises presenting spectro-temporal characteristics similar to speech.
Conversely, a specialized model that was trained on a single speaker but fluctuating noises can show higher performance when evaluated on stationary noises.
Thus, to properly assess the generalization performance, it is necessary to disentangle the effect of the model observing new data from the change in difficulty of the speech enhancement task.

Consider a model $\model$ trained on a dataset $\dtrain$ generated from a condition $\condtrain$, and evaluated on a dataset $\dtest$ generated from a mismatched condition $\condtest$.
To account for the difficulty of the speech enhancement task in $\condtest$, the performance of $\model$ in terms of a given metric is compared to a reference score obtained by a model $\modelref$ trained on a new dataset $\dtrainref$ generated from the test condition $\condtest$.
This is illustrated in Fig.~\ref{fig:gengap}.
The performance of $\modelref$ on $\dtest$ serves as an upper performance limit that can in principle only be reached if we know the test condition beforehand, and allows to separate the performance drop of $\model$ from $\condtrain$ to $\condtest$ into two components.
The first component is related to the change in task difficulty, and corresponds to the performance difference between $\model$ and $\modelref$ in their respective matched configurations.
This is quite intuitive, since if $\condtest$ is indeed more difficult, then the performance of $\modelref$ on $\dtest$ should be lower, even if it was trained on $\condtest$.
The second component is the resulting residual, and is the performance difference between $\model$ and $\modelref$ when both tested on $\dtest$.
We believe this is a more accurate indicator of generalization performance that is less sensitive to the change in difficulty of the task, and we denote it as the generalization gap.
The smaller the generalization gap, the closer the performance of $\model$ to $\modelref$, and the better the generalization to the unseen condition.
A generalization gap equal to zero would indicate maximum generalization; indeed, this would mean $\model$ performs as well as $\modelref$, even though it has not seen $\condtest$ during training, as opposed to $\modelref$.

\subsection{Cross-validation procedure}

To further reduce the influence of a specific choice of databases for training and testing, we cycle through multiple combinations of databases in a cross-validation manner.
We use $M=5$ speech corpora, noise databases and \gls{brir} databases.
For speech, we use TIMIT~\cite{garofolo1993timit}, LibriSpeech (100-hour version)~\cite{panayotov2015librispeech}, WSJ SI-84~\cite{paul1992design}, Clarity~\cite{cox2022clarity} and VCTK~\cite{veaux2013voice}.
For noise, we use TAU~\cite{heittola2019tau}, NOISEX\cite{varga1993noisex}, ICRA~\cite{dreschler2001icra}, DEMAND~\cite{thiemann2013demand} and ARTE~\cite{weisser2019ambisonic}.
Finally for the \glspl{brir}, we use the Surrey~\cite{hummersone2010surrey}, ASH~\cite{shanon2021ash}, BRAS~\cite{brinkmann2021bras}, CATT~\cite{catt_brirs} and AVIL~\cite{mccormack2020higher} databases.
Details about each corpus and database are listed in Tab.~\ref{tab:corpora}.

\ActivateWarningFilters[narrowstyle]
\begin{table}
    \caption{The speech (a), noise (b) and \gls{brir} (c) databases used to generate the acoustic scenes\label{tab:corpora}}
    \centering
    \subfloat[Speech\label{tab:corpora_speech}]{
        \setlength{\tabcolsep}{2.5pt}
        \begin{tabular}{cccccc}
            \toprule
            Corpus
            & TIMIT\,\cite{garofolo1993timit}
            & Libri.\,\cite{panayotov2015librispeech}
            & WSJ\,\cite{paul1992design}
            & Clarity\,\cite{cox2022clarity}
            & VCTK\,\cite{veaux2013voice}
            \\ \midrule
            Speakers & 630 & 251 & 131 & 40 & 110
            \\
            Utterances & 6300 & 28539 & 34738 & 11352 & 44455
            \\
            Hours & 5.4 & 100 & 70 & 9 & 42
            \\ \bottomrule
        \end{tabular}
    }\hfil
    \subfloat[Noise\label{tab:corpora_noise}]{
        \setlength{\tabcolsep}{2.5pt}
        \begin{tabular}{cccccc}
            \toprule
            Database
            & \narrowstyle TAU\,\cite{heittola2019tau}
            & \narrowstyle NOISEX\,\cite{varga1993noisex}
            & \narrowstyle ICRA\,\cite{dreschler2001icra}
            & \narrowstyle DEMAND\,\cite{thiemann2013demand}
            & \narrowstyle ARTE\,\cite{weisser2019ambisonic}
            \\ \midrule
            Noise types & 10 & 15 & 10 & 18 & 13
            \\
            Hours & 40 & 1 & 1 & 1.5 & 0.5
            \\ \bottomrule
        \end{tabular}
    }\hfil
    \subfloat[\gls{brir}\label{tab:corpora_rooms}]{
        \setlength{\tabcolsep}{2.5pt}
        \begin{tabular}{cccccc}
            \toprule
            Database
            & Surrey\,\cite{hummersone2010surrey}
            & ASH\,\cite{shanon2021ash}
            & BRAS\,\cite{brinkmann2021bras}
            & CATT\,\cite{catt_brirs}
            & AVIL\,\cite{mccormack2020higher}
            \\ \midrule
            Rooms & 5 & 40 & 4 & 11 & 4
            \\
            Directions & 37 & 5 to 24 & 45 & 37 & 24
            \\ \bottomrule
        \end{tabular}
    }
    \\[3pt]
    \hspace{-10pt}In (c), ``Directions'' refers to the number of \glspl{brir} per room.
\end{table}
\DeactivateWarningFilters[narrowstyle]

We cannot consider all possible combinations of databases, e.g if we just consider the case with $\nspeech=1$ speech corpus, $\nnoise=1$ noise database and $\nroom=1$ \gls{brir} database, there are $\binom{5}{1}^3=125$ possible combinations.
Similarly for $\nspeech=4$ speech corpora, $\nnoise=4$ noise databases and $\nroom=4$ \gls{brir} databases, there are $\binom{5}{4}^3=125$ possible combinations.
Therefore, we limit the study to the combinations depicted in Tab.~\ref{tab:db_select}.
These combinations were obtained by considering two cases:
\begin{itemize}
    \item $N=1$ or \textit{low diversity} training: we use $N=\nspeech=\nnoise=\nroom=1$.
    The 5 folds were obtained by randomly changing the databases across all dimensions without replacement until all databases are used.
    I.e.\ if we index the databases depicted in Tab.~\ref{tab:corpora} from $j=1$ to $j=5$ for each dimension and use the notations introduced previously, then a model $\modeli$ in fold $i$ is trained on a dataset $\dtraini$ generated from $\strainspeechi=\{\corp{speech}{i}\}$, $\strainnoisei=\{\corp{noise}{i}\}$ and $\strainroomi=\{\corp{room}{i}\}$.
    \item $N=4$ or \textit{high diversity} training: we use $N=\nspeech=\nnoise=\nroom=4$.
    The 5 folds were obtained by changing the discarded database across all dimensions without replacement until all databases are discarded.
    I.e.\ with the same indexation as above, a model $\modeli$ in fold $i$ is trained on a dataset $\dtraini$ generated from $\strainspeechi=\{\corp{speech}{j} \mid j \neq i\}$, $\strainnoisei=\{\corp{noise}{j} \mid j \neq i\}$ and $\strainroomi=\{\corp{room}{j} \mid j \neq i\}$.
\end{itemize}

\begin{table}
\caption{Databases used for training in each fold\label{tab:db_select}}
\newcolumntype{L}[1]{>{\raggedright\let\newline\\\arraybackslash\hspace{0pt}}m{#1}}
\newcolumntype{Z}[1]{>{\centering\let\newline\\\arraybackslash\hspace{0pt}}m{#1}}
\newcolumntype{R}[1]{>{\raggedleft\let\newline\\\arraybackslash\hspace{0pt}}m{#1}}
\setlength\tabcolsep{0pt}
\renewcommand{\arraystretch}{1.3}
\footnotesize
\centering
\subfloat[$N=1$, or \textit{low diversity training}]{
    \centering
    \begin{tabular}{Z{3.3em}O{Z{5.6em}}O{Z{5.6em}}O{Z{5.6em}}O{Z{5.6em}}O{Z{5.6em}}}
        \toprule
        & Fold 1
        & Fold 2
        & Fold 3
        & Fold 4
        & Fold 5
        \\ \midrule
        Speech
        & TIMIT & LibriSpeech & WSJ & Clarity & VCTK
        \\
        Noise
        & TAU & NOISEX & ICRA & DEMAND & ARTE
        \\
        Room
        & Surrey & ASH & BRAS & CATT & AVIL
        \\ \bottomrule
    \end{tabular}
}\hfil
\subfloat[$N=4$, or \textit{high diversity training}]{
    \centering
    \begin{tabular}{Z{3.3em}O{Z{5.6em}}O{Z{5.6em}}O{Z{5.6em}}O{Z{5.6em}}O{Z{5.6em}}}
        \toprule
        & Fold 1
        & Fold 2
        & Fold 3
        & Fold 4
        & Fold 5
        \\ \midrule
        \multirow{2.3}{*}{Speech}
        & \makecell{LibriSpeech\\WSJ\\Clarity\\VCTK}
        & \makecell{TIMIT\\WSJ\\Clarity\\VCTK}
        & \makecell{TIMIT\\LibriSpeech\\Clarity\\VCTK}
        & \makecell{TIMIT\\LibriSpeech\\WSJ\\VCTK}
        & \makecell{TIMIT\\LibriSpeech\\WSJ\\Clarity}
        \\
        \multirow{2.3}{*}{Noise}
        & \makecell{NOISEX\\ICRA\\DEMAND\\ARTE}
        & \makecell{TAU\\ICRA\\DEMAND\\ARTE}
        & \makecell{TAU\\NOISEX\\DEMAND\\ARTE}
        & \makecell{TAU\\NOISEX\\ICRA\\ARTE}
        & \makecell{TAU\\NOISEX\\ICRA\\DEMAND}
        \\
        \multirow{2.3}{*}{Room}
        & \makecell{ASH\\BRAS\\CATT\\AVIL}
        & \makecell{Surrey\\BRAS\\CATT\\AVIL}
        & \makecell{Surrey\\ASH\\CATT\\AVIL}
        & \makecell{Surrey\\ASH\\BRAS\\AVIL}
        & \makecell{Surrey\\ASH\\BRAS\\CATT}
        \\ \bottomrule
    \end{tabular}
}
\end{table}

\begin{figure}
    \footnotesize
    \centering
    \includeinkscape[width=.9\linewidth]{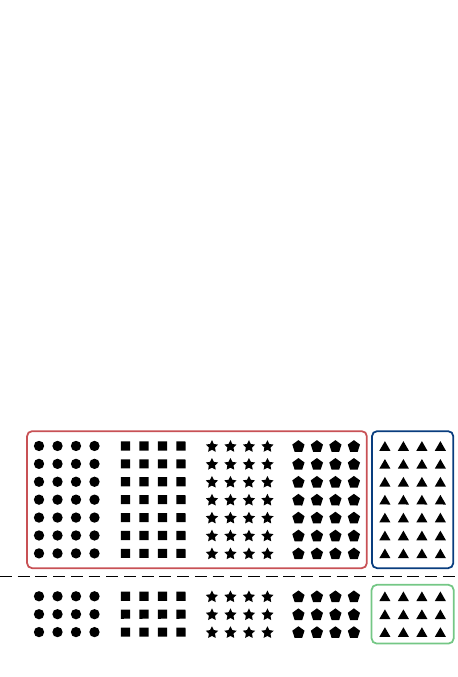}
    \caption{One fold of the cross-validation procedure along the speech dimension.
    Each corpora is split such that no utterance appears during both training and testing.
    Evaluated models are trained on a dataset $\dtrain$ generated from $N=1$ (a) or $N=4$ (b) corpora, and tested on a dataset $\dtest$ generated from the left-out corpora.
    The reference model is trained on a dataset $\dtrain_\text{ref}$ generated from the left-out corpora, and tested on $\dtest$ as well.
    The same procedure is applied to investigate noise and \gls{brir} generalization.}
    \label{fig:crossval}
\end{figure}

For each fold, the trained model is then evaluated on the complementary set of databases along the investigated dimensions.
For example, if a single mismatch along the speech dimension is investigated, the model $\modeli$ in fold $i$ is evaluated on a dataset $\dtesti$ generated from all the speech corpora unused for training $\stestspeechi=\overline{\strainspeechi}$, but the same noise and \gls{brir} databases $\stestnoisei=\strainnoisei$ and $\stestroomi=\strainroomi$.
Moreover, for each fold $i$, a reference model $\modelrefi$ is trained on a dataset $\dtrainrefi$ generated from the test condition as described in Sect.~\ref{sec:ref_model}, i.e.\ from $\stestspeechi$, $\stestnoisei$ and $\stestroomi$, and evaluated on the same test dataset $\dtesti$ as $\modeli$.
For any dimension $d\in\{\text{speech},\text{noise},\text{room}\}$ and for all folds, no speech utterance, noise segment nor \gls{brir} is seen during both training and testing by any model, even when $\straindi$ = $\stestdi$.
For speech, this is ensured by reserving $80\,\%$ of the utterances in each corpus for training and $20\,\%$ for testing.
For noise, each recording is split into two segments; the first segment accounts for $80\,\%$ of the recording and is used for training, while the second segment accounts for the remaining $20\,\%$ and is used for testing.
For \glspl{brir}, every second \gls{brir} of each room is reserved for training and the other half for testing.
Figure~\ref{fig:crossval} illustrates one fold of the cross-validation procedure along the speech dimension and how the evaluated and reference models are trained and tested on different speech utterances.
In total, 40 models are trained for each model architecture.

\subsection{Generalization gap estimation}

The models are evaluated in terms of \gls{pesq}~\cite{recommendation2001perceptual}, \gls{estoi}~\cite{jensen2016algorithm} and \gls{snr} improvements of the average signal across left and right channels.
For all metrics, the direct-sound component of the speech signal is used as the reference signal.
The improvements between the unprocessed input mixture and the enhanced output signal are denoted as $\dpesq$, $\destoi$ and $\dsnr$.

For each metric $\score\in\{\text{$\dpesq$}, \text{$\destoi$}, \text{$\dsnr$}\}$, we define the generalization gap $\gengap$ as the relative difference between the score obtained by the evaluated model and the score obtained by the reference model.
This relative difference is averaged across folds and expressed in percentage,
\begin{equation}
\gengap=100\times\frac{1}{F}\sum_{i=1}^{F}\frac{\scorei-\scorerefi}{\scorerefi},
\label{eq:gen_gap}
\end{equation}
where $\scorei$ and $\scorerefi$ are the scores obtained by the evaluated model $\modeli$ and the reference model $\modelrefi$ respectively in the $i$-th fold.
Since the reference model gives higher scores than the evaluated model, $\gengap$ is negative.
A pseudo-code algorithm describing how the generalization gap is calculated can be found in Appendix~\ref{sec:pseudo_code}.

Note that the proposed generalization gap is a measure of generalization to distributions that are different from the one seen during training.
Common definitions usually assume that the training and testing observations are \gls{iid}~\cite{goodfellow2016deep,murphy2022probabilistic}.
However in speech enhancement, it is common to split the training and testing observations in a non-random way, e.g.\ by speaker.
The training and testing distributions are thus different, since different speakers exhibit different spectro-temporal characteristics, even when they are selected from the same corpus.
Some studies on \gls{ood} generalization or domain generalization avoid any ambiguity by using terms such as ``\gls{iid}/\gls{ood} generalization gap''~\cite{hendrycks2020pretrained,hendrycks2021many}, ``domain gap''~\cite{wei2018person,nam2021reducing} or ``domain generalization gap''~\cite{cha2021swad}.
In the following, the term ``generalization gap'' refers to a measure of \gls{ood} generalization.

\section{Mixture simulation}\label{sec:mixgen}

Each acoustic scene is simulated by convolving a utterance from a target speaker and up to three interfering noise sources with \glspl{brir} corresponding to different and random spatial locations between \SI{-90}{\degree} and \SI{90}{\degree} in the same room.
The convolution outputs are then mixed at a random \gls{snr} uniformly distributed between \SI{-5}{\decibel} and \SI{10}{\decibel}.
The \gls{snr} is defined as the energy ratio between the direct-sound part of the speech signal and the background signal, which consists of the direct-sound and reverberant parts of the noises as well as the reverberant part of the speech.
The direct-sound part of the target include early reflections up to a boundary of \SI{50}{\milli\second}, which was shown to be beneficial for speech intelligibility~\cite{roman2013speech}.
The signals are created by splitting the \glspl{brir} into a direct-sound and a reverberant component using a windowing procedure outlined in~\cite{zahorik2002direct}.

All the models are trained on $30$ hours of simulated noisy and reverberant speech mixtures.
This amount is held constant independently of the combination of databases from which speech utterances, noise recordings and \glspl{brir} are drawn to generate the training mixtures.
This means that when training on more databases, what changes is the diversity of the data, rather than the amount of data seen during training.
This was chosen to allow for a fair comparison between the evaluated and reference models.
The models are then evaluated on $1$ hour of simulated mixtures for each mismatch scenario.

\section{Learning-based speech enhancement systems}
\label{sec:systems}

This section describes the different learning-based systems used to perform the speech enhancement task and whose generalization performance is evaluated in Sect.~\ref{sec:results}.
For all systems, the binaural mixtures are sampled at a rate of \SI{16}{\kilo\hertz} and averaged across left and right ears at the input of the system.
All models are trained on arbitrary-long sequences using a bucket batching strategy with a dynamic batch size as described in~\cite{gonzalez2023batching}.
The batch size is set to \SI{128}{\second} and the number of buckets is set to $10$.

\subsection{FFNN-based system}\label{sec:ffnn}

\glspl{ffnn} have been extensively used for speech enhancement~\cite{wang2012boosting,wang2012cocktail,wang2013towards,healy2013algorithm,wang2014training,xu2015regression,wang2015deep,kolbaek2016speech,chen2016large,may2017robust,bentsen2018benefit,wang2018supervised} and were shown to improve the intelligibility of noisy and reverberant speech mixtures~\cite{chen2016large,may2017robust,bentsen2018benefit}.
The present \gls{ffnn}-based system consists of a \gls{stft} followed by a feature extraction stage and a prediction of the \gls{irm}~\cite{wang2014training} by the \gls{ffnn}.
The \gls{stft} uses a frame length of 512 samples, an overlap of $50\,\%$ and a Hann window.
The squared magnitude of the \gls{stft} coefficients is integrated into a bank of 64 triangular filters evenly spaced on a mel-frequency scale~\cite{davis1980comparison} between \SI{50}{\hertz} and \SI{8}{\kilo\hertz}.
The output from each filter is then log-compressed and used as input feature for the \gls{ffnn}.
Temporal context is included by stacking current feature frames with five previous frames, resulting in a 384-dimensional input.
Only present and previous frames are included to ensure that the system is causal.

The training target of the \gls{ffnn} is the \gls{irm}~\cite{wang2014training}, which is defined for every mel-filter index $m$ and frame index $l$ as
\begin{equation}
    \mathrm{IRM}(m, l) = \sqrt{\frac{\sum_k G_{m, k}|S_{k, l}|^2}{\sum_k G_{m, k}(|S_{k, l}|^2+|N_{k, l}|^2)}},
    \label{eq:irm}
\end{equation}
where $G_{m, k}$ is the gain of the $m$-th mel-filter at frequency bin $k$, and $S_{k, l}$ and $N_{k, l}$ are the \gls{stft} coefficients of the speech direct-sound part and the background signal respectively (see Sect.~\ref{sec:mixgen} for details on how the different signals are defined).
The estimated ratio mask is then extrapolated to the \gls{stft} frequency axis and applied as a gain function before computing the inverse \gls{stft} to reconstruct the enhanced speech signal.

The \gls{ffnn} has one input layer with $384$ nodes, two hidden layers with $1024$ nodes and rectified linear unit activation, and one output layer with $64$ nodes and sigmoid activation.
The model is optimized to minimize the mean squared error of the predicted ratio mask using the Adam algorithm~\cite{kingma2015adam} with a learning rate of $1e^{-4}$ for 100 epochs.
Dropout~\cite{srivastava2014dropout} with a rate of $20\,\%$ is applied after the activation in each hidden layer to prevent overfitting.
The number of layers, their size and the dropout rate were selected after conducting preliminary experiments and are similar to previously used architectures in literature~\cite{wang2012boosting,wang2012cocktail,wang2013towards,healy2013algorithm,wang2014training,xu2015regression,wang2015deep,kolbaek2016speech,chen2016large,may2017robust,bentsen2018benefit}.
The model has $1.51\,\text{M}$ parameters.
During training, each feature dimension is normalized to have zero mean and unit variance, and the same statistics are used to normalize the features during testing.

\subsection{Conv-TasNet}

Conv-TasNet~\cite{luo2019conv} is an end-to-end fully convolutional network designed for single-channel multi-speaker speech separation.
It can be applied for speech enhancement as in~\cite{koyama2020exploring,kinoshita2020improving} by setting the number of separated sources to $K=1$ to estimate the direct-sound component of the speech signal (see Sect.~\ref{sec:mixgen} for details on how the different signals are defined).
We train the network on the \gls{snr} loss instead of the \gls{sisnr}~\cite{leroux2019sdr} loss as in~\cite{kinoshita2020improving} to prevent the network from scaling the enhanced signal.
This was shown to provide superior performance in terms of multiple perceptual metrics~\cite{koyama2020exploring}.
For the encoder/decoder, we use $N=128$ filters with a length of \SI{2}{\milli\second}, i.e.\ $L=32$ at \SI{16}{\kilo\hertz}.
For the separation network, we choose the configuration reporting the best performance in~\cite{luo2019conv}, namely $B=128$, $H=512$, $S_c=128$, $P=3$, $X=8$, $R=3$, and global layer normalization (gLN), making the system non-causal.
The model is trained with the Adam algorithm~\cite{kingma2015adam} for 100 epochs, with a learning rate of $1e^{-3}$ and gradient clipping with a maximum $L_2$-norm of 5.
The model has $4.94\,\text{M}$ parameters (this is slightly lower than the number reported in~\cite{luo2019conv} due to $K=1$).

\subsection{DCCRN}

DCCRN~\cite{hu2020dccrn} is a causal U-Net~\cite{ronneberger2015unet} -based network operating in the \gls{stft} domain.
It combines ideas from DCUNET~\cite{choi2019phase} and CRN~\cite{tan2018convolutional} by using complex-valued convolutions~\cite{trabelsi2018deep} in the encoder/decoder and a \gls{lstm} network~\cite{hochreiter1997long} in the bottleneck layer to capture long-term temporal dependencies from speech.
It predicts a complex mask to reconstruct both the magnitude and phase of the target signal, as opposed to the \gls{ffnn}-based system which predicts a real mask and uses the noisy input phase.
We train it with the \gls{snr} loss instead of the \gls{sisnr} loss similar to Conv-TasNet.
The total number of channels in the encoder (along the real and imaginary axis combined) is set to $[32, 64, 128, 256, 256, 256]$ and we use the ``E'' enhancement strategy.
We use regular batch normalization~\cite{ioffe2015batch} along the real and imaginary axis separately instead of complex batch normalization~\cite{trabelsi2018deep} as we find this greatly improves training speed and stability without reducing performance.
This is in line with recent evidence that the performance gain of DCCRN is not attributed to complex-valued operations~\cite{wu2023rethinking}.
The model is trained during 100 epochs using the Adam optimizer~\cite{kingma2015adam}, with a learning rate of $1e^{-4}$ and gradient clipping with a maximum $L_2$-norm of 5.
The model has $3.67\,\text{M}$ parameters.

\subsection{MANNER}

MANNER~\cite{park2022manner} is an end-to-end multi-view attention network that currently ranks 6\textsuperscript{th} in terms of \gls{pesq} on the Voicebank+DEMAND dataset\footnote{\url{https://paperswithcode.com/sota/speech-enhancement-on-demand}}~\cite{valentini2016speech}.
It presents a U-Net~\cite{ronneberger2015unet} -based architecture, whose blocks combine channel attention~\cite{woo2018cbam} with local and global attention along two signal scales similar to dual-path models~\cite{luo2020dual}.
Local attention is calculated using one-dimensional convolutions, while global attention is captured using self-attention~\cite{vaswani2017attention}.
The model is trained with the same loss as~\cite{defossez2020real}, that is a compound loss combining a time-domain $\mathcal{L}_1$-loss and a multi-resolution \gls{stft} loss~\cite{yamamoto2020parallel}.
We use the ``small'' version of the model provided in~\cite{park2022manner}, which presents multi-attention blocks only in the last encoder and decoder layers, as this greatly reduces inference time and memory usage at a small cost in performance~\cite{park2022manner}.
We use the Adam algorithm~\cite{kingma2015adam} and the OneCycleLR scheduler~\cite{smith2019super} to optimize the model, with a minimum and maximum learning rate of $1e^{-5}$ and $1e^{-2}$ respectively as in the original paper.
The model is trained for 100 epochs and has $21.25\,\text{M}$ parameters.
Despite the high number of parameters, the model achieves similar memory usage, training times and inference times compared to Conv-TasNet and DCCRN thanks to its efficient architecture.

\section{Results}
\label{sec:results}

\subsection{Performance in matched conditions}
\label{sec:res_matched}

Table~\ref{tab:matched_scores} shows the $\dpesq$, $\destoi$ and $\dsnr$ scores obtained by the different models when both trained and tested on the individual folds depicted in Tab.~\ref{tab:db_select}.
It can be seen that the scores vary substantially across folds for all metrics, especially when using $N=1$ database.
This shows that using particular databases for training and testing substantially influences the results.
Moreover, the average scores are lower for $N=4$ compared to $N=1$ for all metrics and for all models, except for DCCRN and MANNER in terms of $\destoi$.
This makes sense since the systems are trained to perform well for a wide range of conditions in the $N=4$ case, while the systems are specialized in the $N=1$ case.
The performance drop between $N=4$ and $N=1$ is the largest for the \gls{ffnn}-based system, while it is the smallest for DCCRN and MANNER.
The difference in this performance drop between the models can be explained by the different model sizes and architectures.
The \gls{ffnn}-based system presents the lowest number of parameters, which might prevent it from modelling the increase in data complexity when increasing the number of databases.
Moreover, the other models present architectures that are more suited for capturing temporal dependencies, especially DCCRN and MANNER with their \gls{lstm} and global self-attention mechanisms respectively.

Conv-TasNet and MANNER achieve the best scores overall, followed by DCCRN and the  \gls{ffnn}-based system.
This can be explained by Conv-TasNet and MANNER being non-causal and utilizing the entire signal at every time step during enhancement; Conv-TasNet uses gLN, while MANNER uses global self-attention.
Conv-TasNet shows the best $\dsnr$ scores on all folds in both $N=1$ and $N=4$, which can be explained by the fact that Conv-TasNet is trained on the \gls{snr} loss.
Conv-TasNet also shows the best scores in terms of the other metrics on average, except in terms of $\dpesq$ for $N=4$, where MANNER shows the best values in all folds.
This suggests Conv-TasNet is suited for speech enhancement in matched conditions, particularly when these are specialized.
However, while the $\destoi$ and $\dsnr$ scores stay high for more complex conditions, the $\dpesq$ scores can drop substantially.

\begin{table*}
\caption{$\dpesq$, $\destoi$ and $\dsnr$ scores obtained for each fold in matched conditions, i.e.\ when using the databases depicted in Tab.~\ref{tab:db_select} during both training and testing\label{tab:matched_scores}}
\vspace{-8pt}
\centering
\sisetup{table-format=2.2,detect-all}
\subfloat[$N=1$\label{tab:matched_scores_a}]{
\centering
\begin{tabular}{cc*{4}{S}}
\toprule
 &  & {FFNN} & {Conv-TasNet} & {DCCRN} & {MANNER} \\
\midrule
\multirow{6.5}{*}{\rotatebox[origin=c]{90}{$\Delta \text{PESQ}$}} & Fold 1 & 0.63 & \bfseries 1.32 & 0.69 & 1.28 \\
 & Fold 2 & 0.33 & 0.57 & 0.41 & \bfseries 0.62 \\
 & Fold 3 & 0.24 & \bfseries 0.68 & 0.29 & 0.49 \\
 & Fold 4 & 0.52 & 0.70 & 0.56 & \bfseries 0.82 \\
 & Fold 5 & 0.44 & 0.70 & 0.54 & \bfseries 0.76 \\
\cmidrule{2-6}
 & Mean & 0.43 & \bfseries 0.80 & 0.50 & 0.79 \\
\midrule
\multirow{6.5}{*}{\rotatebox[origin=c]{90}{$\Delta \text{ESTOI}$}} & Fold 1 & 0.15 & \bfseries 0.28 & 0.15 & 0.25 \\
 & Fold 2 & 0.13 & 0.18 & 0.15 & \bfseries 0.19 \\
 & Fold 3 & 0.17 & \bfseries 0.37 & 0.22 & 0.25 \\
 & Fold 4 & 0.08 & 0.10 & 0.09 & \bfseries 0.11 \\
 & Fold 5 & 0.12 & \bfseries 0.18 & 0.15 & 0.17 \\
\cmidrule{2-6}
 & Mean & 0.13 & \bfseries 0.22 & 0.15 & 0.19 \\
\midrule
\multirow{6.5}{*}{\rotatebox[origin=c]{90}{$\Delta \text{SNR}$}} & Fold 1 & 7.75 & \bfseries 11.01 & 8.93 & 9.56 \\
 & Fold 2 & 5.59 & \bfseries 8.24 & 7.40 & 7.20 \\
 & Fold 3 & 6.59 & \bfseries 11.25 & 8.72 & 7.27 \\
 & Fold 4 & 7.64 & \bfseries 9.10 & 8.42 & 8.21 \\
 & Fold 5 & 3.94 & \bfseries 6.58 & 5.71 & 5.75 \\
\cmidrule{2-6}
 & Mean & 6.30 & \bfseries 9.24 & 7.84 & 7.60 \\
\bottomrule
\end{tabular}
}\hfil
\subfloat[$N=4$\label{tab:matched_scores_b}]{
\centering
\begin{tabular}{cc*{4}{S}}
\toprule
 &  & {FFNN} & {Conv-TasNet} & {DCCRN} & {MANNER} \\
\midrule
\multirow{6.5}{*}{\rotatebox[origin=c]{90}{$\Delta \text{PESQ}$}} & Fold 1 & 0.28 & 0.55 & 0.37 & \bfseries 0.63 \\
 & Fold 2 & 0.30 & 0.66 & 0.42 & \bfseries 0.71 \\
 & Fold 3 & 0.37 & 0.65 & 0.55 & \bfseries 0.78 \\
 & Fold 4 & 0.28 & 0.65 & 0.43 & \bfseries 0.69 \\
 & Fold 5 & 0.30 & 0.62 & 0.47 & \bfseries 0.68 \\
\cmidrule{2-6}
 & Mean & 0.30 & 0.62 & 0.45 & \bfseries 0.70 \\
\midrule
\multirow{6.5}{*}{\rotatebox[origin=c]{90}{$\Delta \text{ESTOI}$}} & Fold 1 & 0.11 & \bfseries 0.20 & 0.15 & 0.19 \\
 & Fold 2 & 0.11 & \bfseries 0.21 & 0.16 & 0.20 \\
 & Fold 3 & 0.10 & 0.16 & 0.13 & \bfseries 0.16 \\
 & Fold 4 & 0.12 & \bfseries 0.22 & 0.17 & 0.21 \\
 & Fold 5 & 0.11 & \bfseries 0.20 & 0.16 & 0.20 \\
\cmidrule{2-6}
 & Mean & 0.11 & \bfseries 0.20 & 0.15 & 0.19 \\
\midrule
\multirow{6.5}{*}{\rotatebox[origin=c]{90}{$\Delta \text{SNR}$}} & Fold 1 & 5.41 & \bfseries 8.63 & 7.16 & 7.04 \\
 & Fold 2 & 5.21 & \bfseries 8.65 & 7.01 & 7.07 \\
 & Fold 3 & 5.32 & \bfseries 7.66 & 6.84 & 6.61 \\
 & Fold 4 & 5.64 & \bfseries 8.80 & 7.56 & 7.25 \\
 & Fold 5 & 5.78 & \bfseries 8.72 & 7.63 & 7.32 \\
\cmidrule{2-6}
 & Mean & 5.47 & \bfseries 8.50 & 7.24 & 7.06 \\
\bottomrule
\end{tabular}
}
\end{table*}

\subsection{Performance in mismatched conditions}\label{sec:res_mismatch}

\subsubsection{Single mismatch}

Figure~\ref{fig:results_single} shows the $\dpesq$, $\destoi$ and $\dsnr$ results obtained by all architectures in either speech, noise or room mismatch conditions, when training with either $N=1$ or $N=4$ databases.
The scores are shown together with the reference model, which we recall was trained on the unseen testing databases for each fold.
The average generalization gap is expressed in percentage and shown with its standard deviation across folds.
As expected, the scores of the reference model are consistently higher compared to the evaluated model.
When going from $N=1$ to $N=4$, the average performance of the evaluated model increases, and the generalization gap decreases, indicating better generalization for all architectures.

\newcommand{\spaceabovesubcaption}{\vspace{-2pt}}
\newcommand{\spaceabovemaincaption}{\vspace{-8pt}}

Speech is the dimension affecting generalization the most, as it shows the largest generalization gaps (e.g.\ $\gdestoi=-68\%$ for Conv-TasNet or $\gdpesq=-50\%$ for MANNER at $N=1$).
It also shows the lowest worst-case scores (e.g.\ $\dsnr=0.82$ and $\destoi=-0.01$ for Conv-TasNet, or $\dpesq=0.09$ for DCCRN at $N=1$).
The \gls{ffnn}-based system is the most robust to unseen speech at $N=1$ as it shows the smallest generalization gaps across all metrics.
Increasing the number of training databases from $N=1$ to $N=4$ substantially improves performance and closes the generalization gaps, but these are still significant, especially for Conv-TasNet (e.g.\ $\gdpesq=-35\%$).
This indicates generalization to unseen speech is a major challenge.
Apart from the \gls{ffnn}, the most robust model to unseen speech at $N=4$ is DCCRN (e.g.\ $\gdpesq=-18\%$), which is in line with evidence from literature suggesting the \gls{lstm} mechanism is suited for generalizing to unseen speakers~\cite{chen2017long}.

The noise dimension is less problematic as reflected by the smaller generalization gaps for all models at both $N=1$ and $N=4$.
High best-case scores are observed even when training on $N=1$ database (e.g.\ $\destoi=0.28$ for Conv-TasNet or $\dpesq=0.86$ for MANNER).
When training on $N=4$ databases, small generalization gaps are observed (e.g.\ $\gdsnr \geq -5\%$ for Conv-TasNet, DCCRN and MANNER).
These results indicate that good generalization to unseen noises is possible by training on a wide range of noise types.
The most robust model to unseen noises at $N=4$ is MANNER, as it presents the smallest generalization gaps for all metrics.
Interestingly, while the \gls{ffnn} is the most robust to unseen noises at $N=1$, it becomes the least robust at $N=4$ for all metrics.
This suggests that recent development in speech enhancement has not only improved performance in matched conditions, but also reduced the generalization gap in noise mismatches.

Similarly, the room dimension is less problematic than the speech dimension, as it shows the smallest generalization gaps at $N=1$ (e.g.\ $\gdpesq=-10\%$ for the \gls{ffnn} or $\gdestoi=-17\%$ for Conv-TasNet).
At $N=4$, the effect of the room dimension is similar to that of the noise dimension, and the most robust model is Conv-TasNet, as it shows the smallest $\gdpesq$ and $\gdestoi$.

\begin{figure*}
  \centering
  \includeinkscape{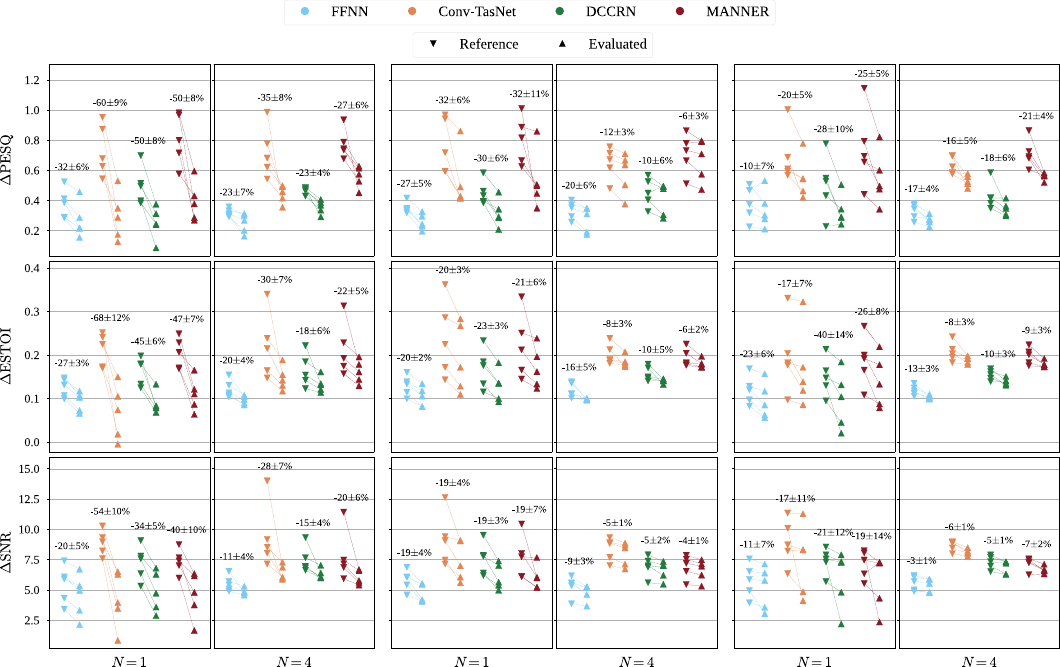}
  \newline
  \begin{minipage}[t]{.04\textwidth}
    \phantom{}
  \end{minipage}%
  \begin{minipage}[t]{.32\textwidth}
    \centering
    \spaceabovesubcaption
    \small (a) Speech%
  \end{minipage}%
  \begin{minipage}[t]{.32\textwidth}
    \centering
    \spaceabovesubcaption
    \small (b) Noise%
  \end{minipage}%
  \begin{minipage}[t]{.32\textwidth}
    \centering
    \spaceabovesubcaption
    \small (c) Room%
  \end{minipage}
  \spaceabovemaincaption
  \caption{Single mismatch experiment results.
  For each fold and architecture, a model is trained on $N=1$ or $N=4$ databases and tested in speech (a), noise (b), or room (c) mismatch.
  A reference model is trained on the test condition.
  The models are evaluated in terms of $\dpesq$, $\destoi$ and $\dsnr$.
  The generalization gap is averaged across folds and expressed in percentage.}
  \label{fig:results_single}
\end{figure*}

\subsubsection{Double mismatch}

\begin{figure*}[!t]
  \centering
  \includeinkscape{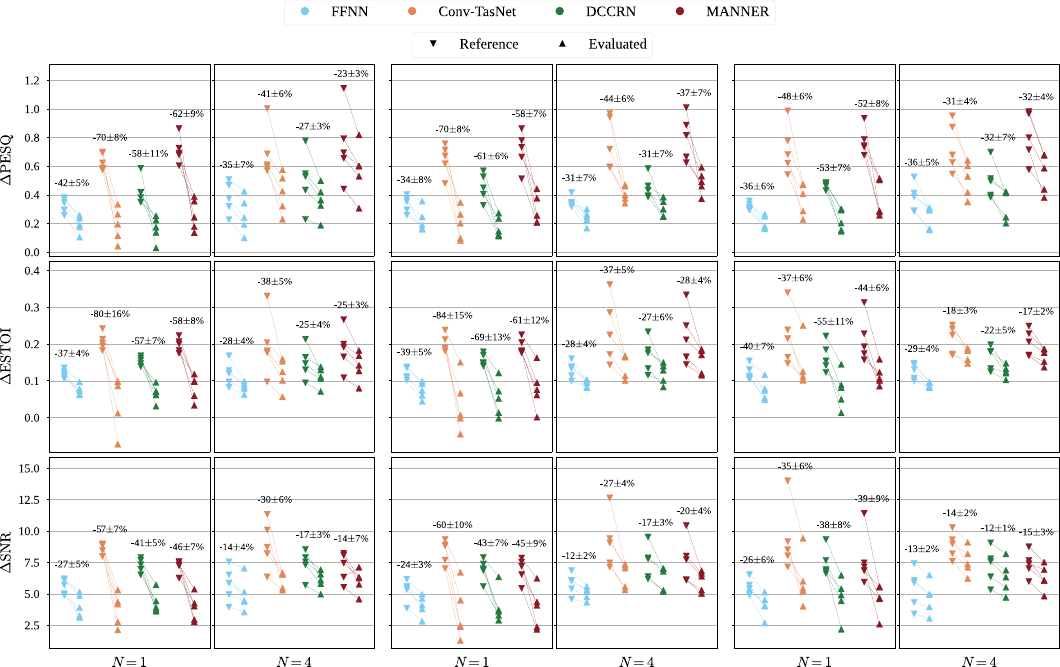}
  \newline
  \begin{minipage}[t]{.04\textwidth}
    \phantom{}
  \end{minipage}%
  \begin{minipage}[t]{.32\textwidth}
    \centering
    \spaceabovesubcaption
    \small (a) Speech + Noise%
  \end{minipage}%
  \begin{minipage}[t]{.32\linewidth}
    \centering
    \spaceabovesubcaption
    \small (b) Speech + Room%
  \end{minipage}%
  \begin{minipage}[t]{.32\linewidth}
    \centering
    \spaceabovesubcaption
    \small (c) Noise + Room%
  \end{minipage}
  \spaceabovemaincaption
  \caption{Double mismatch experiment results.
  For each fold and architecture, a model is trained on $N=1$ or $N=4$ databases and tested in speech and noise (a), speech and room (b), or noise and room (c) mismatch.
  A reference model is trained on the test condition.
  The models are evaluated in terms of $\dpesq$, $\destoi$ and $\dsnr$.
  The generalization gap is averaged across folds and expressed in percentage.}
  \label{fig:results_double}
\end{figure*}

Figure~\ref{fig:results_double} shows the scores obtained by all four models, this time when evaluated in double mismatch conditions, i.e.\ in speech and noise mismatch, speech and room mismatch, or noise and room mismatch.
Once again, the reference scores are consistently higher than the evaluated models as expected.
Moreover, increasing the number of training databases from $N=1$ to $N=4$ closes the generalization gap and improves the scores.
Compared to Fig.~\ref{fig:results_single}, the generalization gaps are overall larger (e.g.\ $\gdestoi=-84\%$ for Conv-TasNet in speech and room mismatch at $N=1$).
The scores are also lower (e.g.\ $\destoi \leq 0 $ in worst-case for Conv-TasNet, DCCRN and MANNER in speech and room mismatch).
Results are consistent with the single mismatch case, in that dimensions that are problematic for a model in the single mismatch case are also problematic in the double mismatch case.
For example, Conv-TasNet, which previously showed to be robust to unseen noises and rooms but sensitive to speech mismatches, is robust in combined noise and room mismatches, but struggles as soon as speech is mismatched.
Similarly, the \gls{ffnn}-based system shows again the smallest generalization gaps across all metrics at $N=1$ (except in terms of $\destoi$ in noise and room mismatch), but the gaps are substantially smaller for the other models when increasing the number of training databases to $N=4$.
DCCRN and MANNER in particular show the smallest gaps in multiple combined mismatches.

\subsubsection{Triple mismatch}

\begin{figure}
  \centering
  \includeinkscape{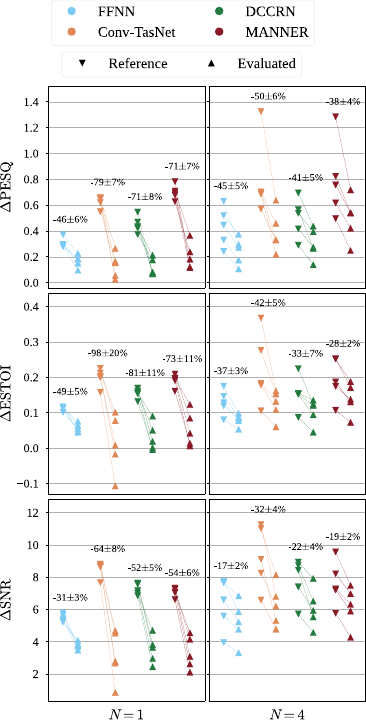}
  \vspace{6pt}
  \caption{Triple mismatch experiment results.
  For each fold and architecture, a model is trained on $N=1$ or $N=4$ databases and tested in speech, noise and room mismatch.
  A reference model is trained on the test condition.
  The models are evaluated in terms of $\dpesq$, $\destoi$ and $\dsnr$.
  The generalization gap is averaged across folds and expressed in percentage.}
  \label{fig:results_triple}
\end{figure}

Figure~\ref{fig:results_triple} shows the results for the triple mismatch case, i.e.\ when speech, noise and room dimensions are all mismatched.
This can be seen as simulating a real-life, in-the-wild scenario, where the target speech, background noise and room differ from the data used during development.
As expected, the scores are even lower and the generalization gaps even larger compared to the double and single mismatch cases.
When using $N=1$ training database, Conv-TasNet, DCCRN and MANNER all show worst-case $\destoi$ scores below $0.01$.
Increasing the number of training databases to $N=4$ substantially reduces the generalization gaps, but these are still significant, especially in terms of $\dpesq$ (from $-50\%$ for Conv-TasNet to $-38\%$ for DCCRN).
This proves that using the same database along all dimensions for training and testing can lead to an overestimation of $\dpesq$ of up to $50\%$ compared to using multiple different databases during training.
Once again, the \gls{ffnn}-based system is the most robust at $N=1$, as per the smallest generalization gaps across all metrics.
At $N=4$, MANNER shows the smallest gaps in terms of $\dpesq$ and $\destoi$, while the \gls{ffnn} still shows the smallest gaps in terms of $\dsnr$.
Conv-TasNet shows the largest gaps across all metrics, whether the number of training databases is $N=1$ or $N=4$.
This suggests Conv-TasNet is not suited for applications where the target speakers, background noise and room are all unknown.

\subsection{Average matched vs.\ mismatched results}

\begin{table*}[!t]
\caption{$\dpesq$, $\destoi$ and $\dsnr$ scores in matched and mismatched conditions, averaged across mismatched dimensions and folds\label{tab:summary}}
\vspace{-8pt}
\footnotesize
\centering
\setlength{\tabcolsep}{5pt}
\sisetup{table-format=2.2,detect-all}
\subfloat[$N=1$\label{tab:summary_a}]{
\centering
\begin{tabular}{cc*{4}{S}}
\toprule
 &  & {FFNN} & {Conv-TasNet} & {DCCRN} & {MANNER} \\
\midrule
\multirow{4}{*}{\rotatebox[origin=c]{90}{$\Delta \text{PESQ}$}} & Match & 0.43 & \bfseries 0.80 & 0.50 & 0.79 \\
 & Single mism. & 0.29 & 0.46 & 0.30 & \bfseries 0.49 \\
 & Double mism. & 0.21 & 0.25 & 0.19 & \bfseries 0.31 \\
 & Triple mism. & 0.17 & 0.13 & 0.12 & \bfseries 0.20 \\
\midrule
\multirow{4}{*}{\rotatebox[origin=c]{90}{$\Delta \text{ESTOI}$}} & Match & 0.13 & \bfseries 0.22 & 0.15 & 0.19 \\
 & Single mism. & 0.10 & \bfseries 0.14 & 0.10 & 0.14 \\
 & Double mism. & 0.07 & 0.07 & 0.06 & \bfseries 0.09 \\
 & Triple mism. & \bfseries 0.06 & 0.01 & 0.03 & 0.05 \\
\midrule
\multirow{4}{*}{\rotatebox[origin=c]{90}{$\Delta \text{SNR}$}} & Match & 6.30 & \bfseries 9.24 & 7.84 & 7.60 \\
 & Single mism. & 4.77 & \bfseries 6.30 & 5.60 & 5.41 \\
 & Double mism. & 4.06 & \bfseries 4.41 & 4.31 & 4.10 \\
 & Triple mism. & \bfseries 3.78 & 3.09 & 3.50 & 3.29 \\
\bottomrule
\end{tabular}
}\hfil
\subfloat[$N=4$\label{tab:summary_b}]{
\centering
\begin{tabular}{cc*{4}{S}}
\toprule
 &  & {FFNN} & {Conv-TasNet} & {DCCRN} & {MANNER} \\
\midrule
\multirow{4}{*}{\rotatebox[origin=c]{90}{$\Delta \text{PESQ}$}} & Match & 0.30 & 0.62 & 0.45 & \bfseries 0.70 \\
 & Single mism. & 0.26 & 0.52 & 0.37 & \bfseries 0.60 \\
 & Double mism. & 0.25 & 0.44 & 0.33 & \bfseries 0.54 \\
 & Triple mism. & 0.24 & 0.39 & 0.30 & \bfseries 0.49 \\
\midrule
\multirow{4}{*}{\rotatebox[origin=c]{90}{$\Delta \text{ESTOI}$}} & Match & 0.11 & \bfseries 0.20 & 0.15 & 0.19 \\
 & Single mism. & 0.10 & 0.17 & 0.14 & \bfseries 0.17 \\
 & Double mism. & 0.09 & 0.14 & 0.12 & \bfseries 0.15 \\
 & Triple mism. & 0.08 & 0.12 & 0.10 & \bfseries 0.14 \\
\midrule
\multirow{4}{*}{\rotatebox[origin=c]{90}{$\Delta \text{SNR}$}} & Match & 5.47 & \bfseries 8.50 & 7.24 & 7.06 \\
 & Single mism. & 5.00 & \bfseries 7.49 & 6.59 & 6.43 \\
 & Double mism. & 4.88 & \bfseries 6.73 & 6.21 & 6.11 \\
 & Triple mism. & 5.19 & \bfseries 6.24 & 6.09 & 6.18 \\
\bottomrule
\end{tabular}
}
\end{table*}

\begin{table*}
\caption{Generalization gaps $G_{\dpesq}$, $G_{\destoi}$ and $G_{\dsnr}$ averaged across mismatched dimensions and folds\label{tab:summary_gg}}
\vspace{-8pt}
\footnotesize
\centering
\setlength{\tabcolsep}{5pt}
\sisetup{table-format=2,detect-all}
\subfloat[$N=1$\label{tab:summary_gg_a}]{
\centering
\begin{tabular}{cc*{4}{S}}
\toprule
 &  & {FFNN} & {Conv-TasNet} & {DCCRN} & {MANNER} \\
\midrule
\multirow{3}{*}{\rotatebox[origin=c]{90}{\scalebox{0.75}{$G_{\Delta \text{PESQ}}$}}} & Single mism. & \bfseries -23\,\% & -37\,\% & -36\,\% & -36\,\% \\
 & Double mism. & \bfseries -37\,\% & -63\,\% & -58\,\% & -57\,\% \\
 & Triple mism. & \bfseries -46\,\% & -79\,\% & -71\,\% & -71\,\% \\
\midrule
\multirow{3}{*}{\rotatebox[origin=c]{90}{\scalebox{0.75}{$G_{\Delta \text{ESTOI}}$}}} & Single mism. & \bfseries -23\,\% & -35\,\% & -36\,\% & -31\,\% \\
 & Double mism. & \bfseries -39\,\% & -67\,\% & -60\,\% & -55\,\% \\
 & Triple mism. & \bfseries -49\,\% & -98\,\% & -81\,\% & -73\,\% \\
\midrule
\multirow{3}{*}{\rotatebox[origin=c]{90}{\scalebox{0.75}{$G_{\Delta \text{SNR}}$}}} & Single mism. & \bfseries -17\,\% & -30\,\% & -25\,\% & -26\,\% \\
 & Double mism. & \bfseries -26\,\% & -51\,\% & -41\,\% & -43\,\% \\
 & Triple mism. & \bfseries -31\,\% & -64\,\% & -52\,\% & -54\,\% \\
\bottomrule
\end{tabular}
}\hfil
\subfloat[$N=4$\label{tab:summary_gg_b}]{
\centering
\begin{tabular}{cc*{4}{S}}
\toprule
 &  & {FFNN} & {Conv-TasNet} & {DCCRN} & {MANNER} \\
\midrule
\multirow{3}{*}{\rotatebox[origin=c]{90}{\scalebox{0.75}{$G_{\Delta \text{PESQ}}$}}} & Single mism. & -20\,\% & -21\,\% & \bfseries -17\,\% & -18\,\% \\
 & Double mism. & -34\,\% & -38\,\% & \bfseries -30\,\% & -30\,\% \\
 & Triple mism. & -45\,\% & -50\,\% & -41\,\% & \bfseries -38\,\% \\
\midrule
\multirow{3}{*}{\rotatebox[origin=c]{90}{\scalebox{0.75}{$G_{\Delta \text{ESTOI}}$}}} & Single mism. & -16\,\% & -16\,\% & \bfseries -12\,\% & -13\,\% \\
 & Double mism. & -28\,\% & -31\,\% & -24\,\% & \bfseries -23\,\% \\
 & Triple mism. & -37\,\% & -42\,\% & -33\,\% & \bfseries -28\,\% \\
\midrule
\multirow{3}{*}{\rotatebox[origin=c]{90}{\scalebox{0.75}{$G_{\Delta \text{SNR}}$}}} & Single mism. & \bfseries -8\,\% & -13\,\% & -8\,\% & -10\,\% \\
 & Double mism. & \bfseries -13\,\% & -24\,\% & -16\,\% & -16\,\% \\
 & Triple mism. & \bfseries -17\,\% & -32\,\% & -22\,\% & -19\,\% \\
\bottomrule
\end{tabular}
}
\end{table*}

To summarize the results, the scores are aggregated across mismatched dimensions to obtain average values in either single, double or triple mismatch, and reported in Tab.~\ref{tab:summary}.
The table also contains the average score in matched conditions.
It can be seen that all systems obtain lower performance in matched conditions when increasing the number of databases.
This is consistent with previous observations and is explained by the models becoming less specialized as the number of training databases increases.
However, the performance increases in mismatched conditions, and the benefit increases as the degree of mismatch increases.
When comparing the different systems, the \gls{ffnn}-based system shows the worst performance in matched conditions across all metrics.
However, the difference with the other models decreases as the degree of mismatch increases, to the point where the \gls{ffnn}-based system shows the best $\destoi$ and $\dsnr$ scores in triple mismatch when training on $N=1$ database.
Conversely, Conv-TasNet provides the best performance in matched conditions across all metrics (except in terms of $\dpesq$ for $N=4$), but shows the largest performance drop in triple mismatch conditions.
When training on $N=4$ databases, the \gls{ffnn}-based system becomes the worst-performing model in triple mismatch across all metrics, while MANNER becomes the best-performing model in terms of $\dpesq$ and $\destoi$.
These results prove that while recent models show strong performance in matched conditions, it is essential to train them with multiple databases in order to obtain good performance in mismatched conditions, or else they can be outperformed by a simple \gls{ffnn}-based system.

Table~\ref{tab:summary_gg} shows the generalization gaps averaged across mismatched dimensions in single, double and triple mismatch.
Consistent to previous observations, the generalization gaps increase for all models as the degree of mismatch increases.
At $N=1$, the \gls{ffnn}-based system shows the smallest gaps for all metrics in all mismatch degrees.
In contrast, Conv-TasNet shows the largest gaps for all metrics in all mismatch degrees (except in terms of $\destoi$ in single mismatch).
When switching to $N=4$, while Conv-TasNet still shows the largest generalization gaps for all metrics in all mismatch degrees, the \gls{ffnn}-based system only shows the smallest gaps in terms of $\dsnr$.
DCCRN and MANNER both show smaller $\gdpesq$ and $\gdestoi$ at $N=4$, with MANNER achieving the smallest gaps in triple mismatch.
This proves that the robustness of recent models can greatly improve if they are trained with multiple databases.

\section{Discussion and limitations}
\label{sec:discussion}

One possible reason for the superior performance in mismatched conditions and the systematically smaller generalization gaps achieved by the \gls{ffnn}-based system at $N=1$ is its model size and architecture.
In addition to presenting the smallest number of parameters, it consists in a feature engineering approach with a non-learnable \gls{stft} magnitude front-end, followed by a mel-filterbank that further reduces the input dimensionality.
The raw \gls{ffnn} output is also a real mask bounded between 0 and 1.
While context frames are fed to the network, the time-dependency of frames is not explicitly modeled due to the \gls{ffnn} architecture.
In contrast, the other models present mechanisms that are more suited for capturing temporal dependencies.
Conv-TasNet and MANNER both use learnable encoder front-ends and their output is unbounded in the time-domain.
DCCRN uses a \gls{stft} front-end, but it leverages an \gls{lstm} network to capture long-term temporal dependencies and predicts a fully-sized complex-valued mask.
For these reasons, these models show superior performance in matched conditions.
However, this also means they are more susceptible to large generalization gaps if the training data is not diverse enough, since the signal characteristics, which were accurately modeled during training, are different in mismatched conditions.
As discussed in Sect.~\ref{sec:res_matched}, this also explains why the \gls{ffnn}-based system shows the largest performance drop in matched conditions from $N=1$ to $N=4$, since it is not able to capture the increase in data complexity as well as the other models.

The plots in Sect.~\ref{sec:res_mismatch} demonstrate the purpose of the reference model, as it can be observed that in almost all configurations, while the performance of models in mismatched conditions varies across folds, the reference model performance varies accordingly.
As explained in \ref{sec:ref_model}, this is due to the test material being more or less challenging in each fold, and introducing the reference model allows to disentangle the change in difficulty of the speech enhancement task from the ability of the system to deal with unseen data.
Moreover, the observed performance spread across folds highlights the risk of over- or under-estimating the performance when simply using one training and one testing databases.
Indeed, as the performance of models in mismatched conditions varies across folds, one could cherry-pick the fold providing the highest score, and erroneously argue for strong generalization as the conditions are effectively mismatched.

One limit of the proposed generalization measure is that the reference model becomes specialized as we increase the number of training databases $N$.
Indeed, since the reference model is trained on the complementary set of training databases used for testing, it is trained on fewer databases as $N$ increases.
This effectively means that while the generalization gap can be reduced, it cannot be closed, and it approaches the difference in performance between a specialized model and a general model.
The generalization gap at $N=4$ can thus be seen as a measure of performance overestimation when using the same database along all dimensions for training and testing.
Note this is only because the total number of considered databases $M=5$ is low, and we thus quickly reach $N=M-1$.
If we were using many more databases, we predict there would be an $N$ for which both the evaluated model and the reference model would be general, and the generalization gap would approach zero.

Alternatively, the reference model can be trained with the databases used to train the evaluated model, in addition to the databases used for testing.
This way, the reference model would see the testing databases during training, but also the databases seen by the evaluated model.
The generalization gap would thus measure how much the reference model benefits from adding the testing databases in the set of training databases.
It becomes obvious that this measure would approach zero as the number of training databases $N$ increases, since the data seen during training by the evaluated model and the reference model would become more similar as $N$ increases.
However, the generalization gap should be a measure of generalization performance, and not a measure of how the evaluated model and the reference model resemble each other.
For this reason, this alternative use of the reference model was not considered.

\section{Conclusion}
\label{sec:conclusions}

When testing a learning-based speech enhancement system in conditions that were not seen during training, the performance of the system can decrease for two reasons: 1) a change in task difficulty and 2) poor model generalization.
To disentangle these two effects, the present work proposes a framework that uses a reference model trained on the test condition, such that the reference model can be used as proxy for the difficulty of the test condition.
The relative difference in performance between the evaluated model and the reference model, termed the generalization gap, is thus a more accurate measure of generalization performance.
Moreover, to reduce the influence of using specific speech, noise and \gls{brir} databases during training and testing, the generalization gap is estimated in a cross-validation fashion by cycling through multiple databases.

This framework was used to evaluate the generalization performance of four learning-based speech enhancement systems, namely a \gls{ffnn}-based system, Conv-TasNet, DCCRN and MANNER.
It was shown that for all models, speech is the acoustic dimension that affects generalization the most compared to noise and room.
Training on a larger number of speech, noise, and \gls{brir} databases improves the results in mismatched conditions and reduces the generalization gaps.
The more recent models show higher performance in matched conditions compared to the \gls{ffnn}-based system.
However, their performance decreases substantially in mismatched conditions when using single speech, noise and \gls{brir} databases during training, to the point where they are outperformed by the \gls{ffnn}-based system in triple mismatch conditions.
Thankfully, this effect is greatly reduced when training on more databases; while Conv-TasNet still shows the largest generalization gaps with $N=4$ databases, DCCRN and MANNER achieve both smaller gaps and better objective metrics compared to the \gls{ffnn}-based system.
These results prove that while recent models show strong performance in matched conditions, it is essential to train them with multiple databases in order to obtain good performance in mismatched conditions.

The proposed framework enables a more accurate assessment of the generalization performance of learning-based speech enhancement systems.
Since the generalization gap is defined as a relative difference between two models with the same architecture, it facilitates a comparison across studies.
With a more accurate measure of generalization potential at hand, promoting generalization of speech enhancement systems becomes easier as this measure can be used as an optimization handle.
The generalization gap can also help to identify the most challenging dimensions for generalization, which can help to design new and more efficient training datasets.

\appendices
\algnewcommand{\LineComment}[1]{\State \(\triangleright\) #1}
\algnewcommand{\InLineComment}[1]{\quad \(\triangleright\) #1}

\section{Pseudo-code for generalization gap evaluation}
\label{sec:pseudo_code}

\vspace{10pt}
\begin{breakablealgorithm}
\caption{Generalization gap evaluation}
\label{alg:gengap_double}
\begin{algorithmic}
\Require $\mathcal{S}_\text{dim}\subseteq\{\text{speech},\text{noise},\text{room}\}$ the set of mismatched dimensions to investigate
\Require 5 different databases $\corpi{d,1}, \ldots, \corpi{d,5}$ for each $d\in\{\text{speech},\text{noise},\text{room}\}$
\Require $N\in\{1,4\}$ the number of databases to use for training
\Require $\model$ the model architecture to investigate
\Require $E$ a performance metric
\Ensure $\gengap$ the generalization gap
\LineComment{Split each database into train and test subsets}
\For{$d \in \{\text{speech},\text{noise},\text{room}\}$}
\For{$j = 1, \ldots, 5$}
    \State Split $\corpi{d,j}$ into $\{\corptraini{d,j}, \corptesti{d,j}\}$
\EndFor
\EndFor
\LineComment{Main loop over folds}
\For{$i = 1, \ldots, 5$}
    \For{$d \in \{\text{speech},\text{noise},\text{room}\}$}
        \LineComment{Define the indices of the training databases as per Tab.~\ref{tab:db_select}}
        \If{$N=1$}\\\hspace{40pt} 
          Let $\mathcal{S}_{i_\text{train}} = \{i\}$
        \ElsIf{$N=4$}\\\hspace{40pt}
          Let $\mathcal{S}_{i_\text{train}} = \overline{\{i\}} = \{j \in \{1, \ldots, 5\}\mid j \neq i\}$
        \EndIf
        \LineComment{Define the indices of the testing databases}
        \If{$d \in \mathcal{S}_\text{dim}$}\\\hspace{40pt}
          Let $\mathcal{S}_{i_\text{test},d} = \overline{\mathcal{S}_{i_\text{train}}} = \{j \in \{1, \ldots, 5\}\mid j \notin \mathcal{S}_{i_\text{train}}\}$ \InLineComment{Remaining indices along mismatched dimensions}
        \Else{}\\\hspace{40pt}
          Let $\mathcal{S}_{i_\text{test},d} = \mathcal{S}_{i_\text{train}}$ \InLineComment{Same indices along matching dimensions}
        \EndIf
    \EndFor
    \State Generate $\dtraini$ using $\{\corptraini{d,j} \mid j\in\mathcal{S}_{i_\text{train}}\}$ for $d\in\{\text{speech},\text{noise},\text{room}\}$
    \State Generate $\dtrainrefi$ using $\{\corptraini{d,j} \mid j\in\mathcal{S}_{i_\text{test},d}\}$ for $d\in\{\text{speech},\text{noise},\text{room}\}$
    \State Generate $\dtesti$ using $\{\corptesti{d,j} \mid j\in\mathcal{S}_{i_\text{test},d}\}$ for $d\in\{\text{speech},\text{noise},\text{room}\}$
    \State Initialize $\modeli$ with architeture $\model$
    \State Train $\modeli$ on $\dtraini$
    \State Let $\scorei$ the score of $\modeli$ when evaluated on $\dtesti$ in terms of the performance metric $E$
    \State Initialize $\modelrefi$ with architeture $\model$
    \State Train $\modelrefi$ on $\dtrainrefi$
    \State Let $\scorerefi$ the score of $\modelrefi$ when evaluated on $\dtesti$ in terms of the performance metric $E$
\EndFor
\State Calculate $\gengap=100\times\frac{1}{F}\sum_{i=1}^{F}\frac{\scorei-\scorerefi}{\scorerefi}$ \InLineComment{Average relative difference across folds}
\end{algorithmic}
\end{breakablealgorithm}
\vspace{5pt}

\bibliographystyle{IEEEtran}
\bibliography{IEEEabrv, abbrv, refs}

\end{document}